\documentclass[10pt]{article}

\usepackage{fullpage}

\usepackage{amsmath}
\usepackage{url}

\begin{document}

\title{Misfortunes of a mathematicians' trio using\\Computer Algebra Systems:\\Can we trust?\thanks{Partially supported by grants MTM2012-36732-C03-02, MTM2012-36732-C03-03 (Ministerio de Econom\'{\i}a y Competitividad), FQM-262, FQM-4643, FQM-7276 (Junta de Andaluc\'{\i}a) and Feder Funds (European Union).}}

\author{Antonio J. Dur\'an\textsuperscript{1}, Mario P\'erez\textsuperscript{2} and Juan L. Varona\textsuperscript{3}\\[6pt]
\small\textsuperscript{1}Dpto.\ de An\'alisis Matem\'atico,
Universidad de Sevilla,
41080 Sevilla, Spain\\[-3pt]
\small email: \url{duran@us.es}\\[2pt]
\small\textsuperscript{2}Dpto.\ de Matem\'aticas,
Universidad de Zaragoza,
50009 Zaragoza, Spain\\[-3pt]
\small email: \url{mperez@unizar.es}\\[2pt]
\small\textsuperscript{3}Dpto.\ de Matem\'aticas y Computaci\'on,
Universidad de La Rioja, 
26004 Logro\~no, Spain\\[-3pt]
\small email: \url{jvarona@unirioja.es}}

\date{October 15, 2013}

\maketitle


\begin{abstract}
Computer algebra systems are a great help for mathematical research but sometimes
unexpected errors in the software can also badly affect it. As an example, we show how we have detected an error of Mathematica computing determinants of matrices of integer numbers: not only it computes the determinants wrongly, but also it produces different results if one evaluates the same determinant twice.

\smallskip

\end{abstract}

\section*{Introduction}

Nowadays mathematicians often use computer algebra systems as an assistant in their mathematical research. Mathematicians have the ideas, and tedious computations are left to the computer. Everybody ``knows'' that computers perform this work better than persons. But, of course, we must trust in the results derived by the powerful computer algebra systems that we use.

Currently we are using Mathematica to find examples and counterexamples of some mathematical results that we are working out, with the aim of finding the correct hypothesis and later to build a mathematical proof. Our goal was to improve some results by Karlin and Szeg\H{o} \cite{KS} related to orthogonal polynomials on the real line. Details are not important, and this is just an example of the use of a computer algebra system by a typical mathematician in research, but let us explain it briefly; it is not necessary to completely understand it, just to see that it was a typical mathematical research with computer algebra as a tool.

Our starting point is a discrete positive measure on the real line $\mu = \sum_{n\ge0} M_n \delta_{a_n}$ (where $\delta_a$ denotes a Dirac delta in~$a$, and $a_n<a_{n+1}$), having a sequence of orthogonal polynomials $\{P_n\}_{n\ge0}$ (where $P_n$ has degree $n$ and positive leading coefficient). Karlin and Szeg\H{o} considered in 1961 (see~\cite{KS})
the $l\times l$ Casorati determinants
\begin{equation}
\label{eq:det}
\det
\begin{pmatrix}
   P_{n}(a_k) & P_{n}(a_{k+1})& \dots & P_{n}(a_{k+l-1}) \\
   P_{n+1}(a_k) & P_{n+1}(a_{k+1})& \dots & P_{n+1}(a_{k+l-1}) \\
   \vdots & \vdots& \vdots & \vdots \\
   P_{n+l-1}(a_k) & P_{n+l-1}(a_{k+1})& \dots & P_{n+l-1}(a_{k+l-1})
\end{pmatrix},\quad n,k\ge 0 .
\end{equation}
They proved that under the assumption that $l$ is even, these determinants are positive for all nonnegative integers $n, k$. Notice that the set of indices $\{n,n+1,\ldots , n+l-1\}$ for the polynomials $P_n$ is formed by consecutive nonnegative integers. We are working out an extension of this remarkable result for more general sets of indices $F$ than those formed
by consecutive nonnegative integers. We have some conjectures which we want to prove or disprove.

We do not have any proof for our conjectures yet and, as far as we can see, this task seems to be very difficult. On the other hand, and just in case, with the help of our computer algebra system, we have been trying to find a counterexample for our conjectures. Eventually, these experiments can also somehow enlighten the problem.

We have then proceeded to construct orthogonal polynomials with respect to discrete positive measures (with a finite number of Dirac deltas; actually this is not a restriction for our conjectures) by mean of its moments. Fixing a set of indices $F=\{f_1,\ldots , f_{l}\}$, $f_i<f_{i+1}$, for the polynomials $P_n$, we have finally evaluated
the determinants
\begin{equation}
\label{eq:det2}
\det
\begin{pmatrix}
   P_{f_1}(a_k) & P_{f_1}(a_{k+1})& \dots & P_{f_1}(a_{k+l}) \\
   P_{f_2}(a_k) & P_{f_2}(a_{k+1})& \dots & P_{f_2}(a_{k+l}) \\
   \vdots & \vdots& \vdots & \vdots \\
   P_{f_{l}}(a_k) & P_{f_{l}}(a_{k+1})& \dots & P_{f_{l}}(a_{k+l})
\end{pmatrix}
\end{equation}
for a large range of $k$ looking for some negative value.

To avoid the usual problems of float numbers and their algorithms (rounding, truncating, instability), we construct all our examples with integers: by taking integers as the values of $a_n$ and the mass points $M_n$ of the measure, and a suitable normalization of the orthogonal polynomials $P_n$, only integer numbers are involved in~\eqref{eq:det2}. So the computations should be a routine for a computer algebra system, and one could completely trust in the results. We have also introduced random parameters (also integers, of course) to easily perform many experiments.

With the help of Mathematica, one of us found some counterexamples to our conjectures. Fortunately, another of us was using Maple, and when checking those supposed counterexamples he found that they were not counterexamples at all. After revising our algorithms from the scratch, we conclude that either the computations performed with Mathematica or the computations performed with Maple had to be wrong. Things started to be clear when the one of us working with Mathematica found also some counterexamples for the above mentioned result by Karlin and Szeg\H{o} for the case~\eqref{eq:det} and, even more important, his algorithm provided different outputs with the same inputs. Our conclusion was that Mathematica should be computing wrongly. However, our mathematical problem (and our algorithm) was too complicate to convince anybody that Mathematica was making mistakes when managing integers.

\section*{Isolating the error}

Trying to isolate the computational problem, we finally identified that, in some circumstances, Mathematica makes strange mistakes computing determinants whose entries are big integers.
Errors do not occur always, only in some cases. Even worst, over the \emph{same} matrix, the determinant function can get different values!
It resembles the well-known Pentium division bug discovered by Thomas Nicely in 1994, that only affected to several kinds of numbers. But perhaps Mathematica is a black box darker that the internals of a micro, so it is difficult to try to understand what kind of numbers are affected by the Mathematica bug that we are describing.

Instead, we have devised a method to easily generate matrices of big integer numbers that can be represented in a paper and, moreover, whose determinants are clearly erroneously evaluated by Mathematica. As the error not always arises, we show a random procedure to generate these matrices.
Firstly, we generate a random $14\times14$ matrix whose entries are integer numbers between $-99$ and $99$, that is
\begin{verbatim}
  basicMatrix = Table[Table[RandomInteger[{-99, 99}], {i, 1, 14}], {j, 1, 14}]
\end{verbatim}
To have big integers, we multiply every column by $10$ raised to some power, which is the same that multiplying by a diagonal matrix; in particular, we will take
\begin{verbatim}
  powersMatrix = DiagonalMatrix[{10^123, 10^152, 10^185, 10^220, 10^397,
    10^449, 10^503, 10^563, 10^979, 10^1059, 10^1143, 10^1229, 10^1319, 10^1412}]
\end{verbatim}
Finally, to avoid having only integers finishing in many zeroes, we sum a small random matrix given by
\begin{verbatim}
  smallMatrix = Table[Table[RandomInteger[{-999, 999}], {i, 1, 14}], {j, 1, 14}]
\end{verbatim}
Then, we take
\begin{verbatim}
  bigMatrix = basicMatrix.powersMatrix+smallMatrix
\end{verbatim}
(in Mathematica notation, the point \texttt{.} is used to denote the product of matrices). Now we compute the determinant twice:
\begin{verbatim}
  a = Det[bigMatrix];
  b = Det[bigMatrix];
\end{verbatim}
Surprisingly, we very often see that \texttt{a} and \texttt{b} take different values! We can easily see it by checking \texttt{a==b}, that very often generates the answer \texttt{False}, or by visually comparing their numerical approximations \texttt{N[a]} and \texttt{N[b]}.

Let us see a particular example of a real execution of these procedures: with\par\nobreak
\begin{scriptsize}
\[
  \texttt{basicMatrix} =
  \left(
\begin{array}{rrrrrrrrrrrrrr}
 -32 & 69 & 89 & -60 & -83 & -22 & -14 & -58 & 85 & 56 & -65 & -30 & -86 & -9 \\
 6 & 99 & 11 & 57 & 47 & -42 & -48 & -65 & 25 & 50 & -70 & -3 & -90 & 31 \\
 78 & 38 & 12 & 64 & -67 & -4 & -52 & -65 & 19 & 71 & 38 & -17 & 51 & -3 \\
 -93 & 30 & 89 & 22 & 13 & 48 & -73 & 93 & 11 & -97 & -49 & 61 & -25 & -4 \\
 54 & -22 & 54 & -53 & -52 & 64 & 19 & 1 & 81 & -72 & -11 & 50 & 0 & -81 \\
 65 & -58 & 3 & 57 & 19 & 77 & 76 & -57 & -80 & 22 & 93 & -85 & 67 & 58 \\
 29 & -58 & 47 & 87 & 3 & -6 & -81 & 5 & 98 & 86 & -98 & 51 & -62 & -66 \\
 93 & -77 & 16 & -64 & 48 & 84 & 97 & 75 & 89 & 63 & 34 & -98 & -94 & 19 \\
 45 & -99 & 3 & -57 & 32 & 60 & 74 & 4 & 69 & 98 & -40 & -69 & -28 & -26 \\
 -13 & 51 & -99 & -2 & 48 & 71 & -81 & -32 & 78 & 27 & -28 & -22 & 22 & 94 \\
 11 & 72 & -74 & 86 & 79 & -58 & -89 & 80 & 70 & 55 & -49 & 51 & -42 & 66 \\
 -72 & 53 & 49 & -46 & 17 & -22 & -48 & -40 & -28 & -85 & 88 & -30 & 74 & 32 \\
 -92 & -22 & -90 & 67 & -25 & -28 & -91 & -8 & 32 & -41 & 10 & 6 & 85 & 21 \\
 47 & -73 & -30 & -60 & 99 & 9 & -86 & -70 & 84 & 55 & 19 & 69 & 11 & -84 \\
\end{array}
\right)
\]
\end{scriptsize}%
and
\begin{scriptsize}
\[
  \texttt{smallMatrix} =
\left(
\begin{array}{rrrrrrrrrrrrrr}
 528 & 853 & -547 & -323 & 393 & -916 & -11 & -976 & 279 & -665 & 906 & -277 &
   103 & -485 \\
 878 & 910 & -306 & -260 & 575 & -765 & -32 & 94 & 254 & 276 & -156 & 625 & -8 &
   -566 \\
 -357 & 451 & -475 & 327 & -84 & 237 & 647 & 505 & -137 & 363 & -808 & 332 & 222
   & -998 \\
 -76 & 26 & -778 & 505 & 942 & -561 & -350 & 698 & -532 & -507 & -78 & -758 &
   346 & -545 \\
 -358 & 18 & -229 & -880 & -955 & -346 & 550 & -958 & 867 & -541 & -962 & 646 &
   932 & 168 \\
 192 & 233 & 620 & 955 & -877 & 281 & 357 & -226 & -820 & 513 & -882 & 536 &
   -237 & 877 \\
 -234 & -71 & -831 & 880 & -135 & -249 & -427 & 737 & 664 & 298 & -552 & -1 &
   -712 & -691 \\
 80 & 748 & 684 & 332 & 730 & -111 & -643 & 102 & -242 & -82 & -28 & 585 & 207 &
   -986 \\
 967 & 1 & -494 & 633 & 891 & -907 & -586 & 129 & 688 & 150 & -501 & -298 & 704
   & -68 \\
 406 & -944 & -533 & -827 & 615 & 907 & -443 & -350 & 700 & -878 & 706 & 1 & 800
   & 120 \\
 33 & -328 & -543 & 583 & -443 & -635 & 904 & -745 & -398 & -110 & 751 & 660 &
   474 & 255 \\
 -537 & -311 & 829 & 28 & 175 & 182 & -930 & 258 & -808 & -399 & -43 & -68 &
   -553 & 421 \\
 -373 & -447 & -252 & -619 & -418 & 764 & 994 & -543 & -37 & -845 & 30 & -704 &
   147 & -534 \\
 638 & -33 & 932 & -335 & -75 & -676 & -934 & 239 & 210 & 665 & 414 & -803 & 564
   & -805
\end{array}
\right)
\]
\end{scriptsize}%
we have got $\texttt{N[a]} = -3.263388173990166 \cdot 10^{9768}$ and $\texttt{N[b]} = -8.158470434975415 \cdot 10^{9768}$
(if you execute the same program more times, you can get different values).
Actually, none of the above mentioned values is correct, because the value of the determinant of \texttt{bigMatrix} is,
approximately, $1.95124219131987 \cdot 10^{9762}$ (obtained with both Maple and Sage).

We have found this erroneous behavior from Mathematica version~8 (released on November~15, 2010) until the current version 9.0.1,
both under Mac and Windows.
It seems that it does not affect to versions~6 and~7, at least in the same range of numbers.

We have reported the bug on October~7, 2013 (reference CASE:303438), and we have received a kind answer from Wolfram Research Inc.:
\begin{quote}
%
%
It does appear there is a serious mistake on the determinant operation you mentioned. I have forwarded an incident report to our developers with the information you provided.

We are always interested in improving Mathematica, and I want to thank you for bringing this issue to our attention. If you run into any other behavior problems, or have any additional questions, please don't hesitate to contact us.
\end{quote}

We hope that this can be fixed in a near future. However, we have received similar messages in the past, when one of us reported other bugs (for instance, but not only, some of those explained in~\cite{CV}), but without solving them in next releases. In any case, all we can do is wait.

\section*{Conclusions}

We have been using Mathematica as a tool for our mathematical research. All our computations with Mathematica were symbolic, involving only integers (big integers, about 10 thousand digits) and polynomials (with degree 60 at most), so no numerical rounding or instability can arise in them, and we completely trusted the results generated by Mathematica. However, we have obtained completely erroneous results. Perhaps someone can think that this was an esoteric error, without real weightiness, because big integers do not appear in the real life. But this is not the case, because big integers are commonly used, for instance, in cryptography, so it should work without errors. Then, how can we trust computer algebra systems?

We know that it is very difficult to avoid errors in non-trivial programs, so a big effort is necessary to check them. The commercial computer algebra systems are black boxes and their algorithms are opaque to the users (of course, also the source code), and certainly this does not contribute to avoid errors. It makes difficult to apply modern techniques of software verification to this kind of systems (as an example of verification in the context of an open source computer algebra systems, see~\cite{LMRR}). Moreover, known bugs of computer algebra systems should be available to the users; this is usual in free software, but an anathema for commercial packages.

Once stated that criticism, let us stress that software systems have been very useful to help mathematicians.
Some well-known challenges were the proof of the four color problem by Kenneth Appel and Wolfgang Haken \cite{AH} and the Kepler conjecture by Thomas Hales \cite{H}, and less known is the recent success of the mathematical software Kenzo detecting an error in a published mathematical theorem (see~\cite{RR}).
Let us hope that software bugs do not prevent us to continue this fruitful line in the future.

\end{document}